# Effect of quenching medium on short-range order in Fe-rich Fe-Cr alloys

Stanisław M. Dubiel[1*], Jakub Cieślak[1] and Jan Żukrowski[1,2]

[1]AGH University of Science and Technology, Faculty of Physics and Applied Computer Science, al. A. Mickiewicza 30, 30-059 Kraków, Poland, [2]AGH University of Science and Technology, Academic Center for Materials and Nanotechnology, al. A. Mickiewicza 30, 30-059 Kraków, Poland

## Abstract

Effect of a quenching medium (water, liquid nitrogen and block of brass) on a short-range ordering in $Fe_{100-x}Cr_x$ (x ≤ 19) alloys was studied with the Mössbauer spectroscopy. The distribution of Cr atoms was expressed in terms of the Cowley-Warren short-range order (SRO) parameters: $<\alpha_1>$ for the first neighbor-shell, $<\alpha_2>$ for the second neighbor-shell and $<\alpha_{12}>$ for both neighbor-shells. It was revealed that none of the quenching media resulted in a random distribution of atoms, yet the degree of randomness was the highest for the samples quenched onto the block of brass. The quenching into water and liquid nitrogen caused a partial oxidation of samples' surface accompanied by a chromium depletion of the bulk. Quantitative analysis of various phases in the studied samples both in their bulk as well as in pre surface zones was carried out.

[*] **Corresponding author: Stanislaw.Dubiel@fis.agh.edu.pl (S. M. Dubiel)**

## 1. Introduction

$Fe_{100-x}Cr_x$ alloys have been subject of intensive studies due to both their interesting physical properties [1] as well as to their industrial importance [2]. Consequently, they have been regarded and treated as model alloys: by physicists for testing various models and theories aimed at explanation and understanding of these properties, and by metallurgists to better understand and improve materials properties. A distribution of Cr atoms has been of a particular interest, especially around a critical concentration, $x \approx 10$, above which the alloys become stainless. The existence of such critical content was found experimentally from diffuse-neutron-diffraction experiments [3,4], according to which the Cowley-Warren short-range order (SRO) parameter, $\alpha_{12}$, average over the first-two neighbor shells, 1NN-2NN, changed its sign from negative ($x \leq \sim 10$) – indicative of repulsion between Cr atoms – to positive ($x \geq \sim 10$) – indicative of attraction (clustering) between Cr atoms. This finding was also confirmed by other techniques viz. the Mössbauer spectroscopy (MS) [5] and the synchrotron X-ray desorption technique [6]. The inversion of the SRO parameter in the Fe-Cr alloys was first predicted theoretically by *ab initio* calculations [7]. Its existence was also confirmed by calculations of the mixing entropy [8] as well as of the pair potentials applying different approaches [9-16]. It must be, however, realized that according to the atomistic Monte Carlo simulations [15], the inversion in the SRO parameter can be reproduced in thermodynamic equilibrium only when contributions of the Fe-rich ($\alpha$) and the Cr-rich ($\alpha$') phases are taken into account. In other words, the phase separation into $\alpha$ and $\alpha$' phases is a necessary condition for the inversion to occur. Otherwise, the SRO parameter has a minimum whose position and depth depend on the annealing temperature. These simulations clearly demonstrated that the metallurgical state of samples plays a crucial role in the actual distribution of Cr atoms, hence in the values of the SRO parameters. Indeed, our recent study on a series of Fe-Cr alloys carried out with MS gave a sound evidence of this behavior [17]. It has been a general believe and practice that an isothermal annealing at high temperature followed by a rapid cooling results in a homogenous distribution of atoms. However, recent experiments gave evidence that this was not the case [19,20]. Theoretical calculations predict that magnetism has a crucial effect on the phase diagram of the Fe-Cr alloy system and related phenomena [6,21-27]. Short-range order magnetic correlations that exist above the Curie temperature are very important [25,26], and they may be responsible for a non-random distribution of atoms at high temperatures at which the annealing process is performed. Consequently, the following quenching freezes this non-random distribution. Here, the effect of a solution treatment into different quenching media on the actual distribution of Cr atoms in a number of Fe-rich the Fe-Cr alloys is reported.

**2. Samples and spectra measurements**

Master alloys of $Fe_{100-x}Cr_x$ (x≤19) were obtained by melting elemental iron and chromium in an arc furnace under a protective atmosphere of argon. Obtained ingots were cold rolled to the final thickness of ~20-30 μm. Finally, the samples in form of discs ~13mm in diameter and ~10mg in mass were cut from the foils. They underwent three different heat treatments denoted as T1, T2, and T3:

• T1 – annealing at 800K for 3 h in argon, followed by quenching into liquid nitrogen (80K).

• T2 – annealing at 800K for 3 h in argon, followed by quenching into water (290K)

• T3 – annealing at 800K for 3 h in vacuum, followed by quenching onto a block of brass (295K)

Chemical composition of the investigated samples and specification of the applied quenching for particular samples is given in Table 1.

Table 1

List of investigated $Fe_{100-x}Cr_x$ samples with different quenching medium: T1 stands for liquid nitrogen, T2 for water, and T3 for brass. The model EFDA samples [17] are marked with asterisk.

| No | 1 | 2 | 3 | 4 | 5[*] | 6 | 7 | 8 | 9 | 10[*] | 11 | 12 | 13 | 14[*] | 15 | 16 |
|---|---|---|---|---|---|---|---|---|---|---|---|---|---|---|---|---|
| x(at%) | 2.2 | 3.3 | 3.9 | 4.85 | 5.8 | 6.4 | 7.85 | 8.5 | 10.25 | 10.75 | 12.3 | 14.15 | 14.9 | 15.15 | 17 | 19 |
| T1 | + | + | + | + |  | + | + |  | + |  | + | + | + |  |  |  |
| T2 | + | + | + | + |  | + |  |  |  |  |  |  |  |  |  |  |
| T3 | + | + | + |  | + | + | + | + |  | + |  | + |  | + | + | + |

The samples were investigated by means of the Mössbauer spectroscopy that has proved to be a very useful tool in the study of issues related to a distribution of atoms in Fe-alloys, and, in particular, Fe-Cr ones e. g. [17-20,28,29]. [57]Fe Mössbauer spectra we recorded at room temperature using a [57]Co/Rh source for the 14.4 keV gamma rays and a standard spectrometer with a drive operating in a sinusoidal mode. They were recorded both in a transmission (TRANS) as well as in a conversion electrons (CEMS) modes. For the former a proportional counter was used to count the gamma rays while the conversion electrons were registered with a proportional gas flow counter with a He/methane mixture as counting gas.

**3. Spectra analysis**

The spectra characteristic of a single metallic phase, Fe-Cr, were analyzed in terms of the two-shell model (1NN-2NN), aimed at determining short-range-order (SRO) parameters. An effect of the presence of Cr atoms in the 1NN-2NN vicinity of the [57]Fe probe nuclei on the hyperfine field ($H$) and on the isomer shift ($IS$) was assumed to be additive i.e. $X(m,n) = X(0,0) + m\Delta X1 + n\Delta X2$, where $X = H$ or $IS$, and $\Delta Xi$ is a change of $H$ or $IS$ due to one Cr atom situated in 1NN (i=1) or on 2NN (i=2). Twenty five most significant atomic configurations, *(m,n)*, taken into account were chosen based on the binomial distribution. However, their probabilities, $P(m,n)$, were treated as free parameters (their starting values were those from the binomial distribution). All the spectra were fitted simultaneously with a least-squares method assuming the same values of $\Delta Xi$'s. All other spectral parameters like $X(0,0)$, line widths of individual sextets $G1$, $G2$ and $G3$ and their relative intensities (Clebsch-Gordan coefficients) $C2$ and $C3$ were treated as free ($C1$=1). Very good fits (in terms of $\chi^2$) were obtained with the following values of the spectral parameters: $\Delta H1$= -30.5 kOe, $\Delta H2$= -19.5 kOe, $\Delta IS1$= -0.020 mm/s, $\Delta IS2$= -0.007 mm/s, $G1$=0.28(2) mm/s, $G2$=0.30(2) mm/s, $G3$=0.32(2) mm/s, $C2$=2.2(4), $C3$=2.5(1). More details on the applied fitting procedure can be found elsewhere [18,19]. Using the $P(m,n)$-values, the average numbers of Cr atoms in *1NN*, $<n_1>$, in *2NN*, $<n_2>$ neighbor shells were calculated. Their knowledge permitted determination of the Warren-Cowley SRO parameters based on the following equation:

$$<\alpha_k> = 1 - \frac{<n_k>}{<n_r>} \qquad (1)$$

Where $k$=1, 2, 12 for *1NN*, *2NN* and *1NN+2NN*, respectively. The average number of Cr-atoms for a random distribution is indicated by $<n_r>$.

Mössbauer spectra recorded in the CEMS mode were found to be multicomponent in some cases and consisted of up to four phases. Three of them were identified as the following oxides: Cr-substituted magnetite, $Fe_{3-y}Cr_yO_4$, hematite, $Fe_{2-y}Cr_yO_3$, wüstite, $Fe_{1-y}Cr_yO$, and the fourth one as metallic Fe-Cr. Their spectra were analyzed as follows: two sextets for the magnetite, one sextet for the hematite, a distribution of the isomer shift for the wüstite ($Fe_{1-\delta}O_\delta$), and a distribution of the hyperfine field for Fe-Cr. By integrating the distributions average values of the isomer shift, $<IS>$, and that of the hyperfine field, $<H>$, were obtained.

All spectral parameters obtained from the best-fit refinement of the spectra are displayed in Tables 2, 3 and 4.

## 4. Results

### 4.1. Treatment T1

Examples of the spectra recorded for this treatment are presented in Fig. 1 and the best-fit spectral parameters are displayed in Table 2.

One can see that there is a significant difference between the two types of the spectra for the samples with $x \leq 10.25$. Namely, those recorded in the CEMS mode i.e. containing an information from a pre surface zone (its thickness $\leq 0.3$ μm) contain, in addition to a metallic phase (Fe-Cr), up to three different Fe-bearing oxide phases which have been identified as $Fe_{3-y}Cr_yO_4$, hematite and wüstite. Based on the published data concerning the Mössbauer spectroscopic study of Cr-substituted magnetite [30] one can conclude that the concentration of Cr in $Fe_{3-y}Cr_yO_4$ of our samples, $y$, quenched into LN ranges between 0.25 and 0.375. In the $Fe_{89.25}Cr_{10.25}$ sample there was found only wüstite while the $Fe_{83}Cr_{17}$ sample was purely metallic. A relative amount of the Fe-bearing oxide phases, $A$, determined from the relative spectral area, decreases steeply with the increase of chromium concentration from ~80% for $x$=2.2 to ~25% for $x$=10.25. This behavior reflects the known fact that

Table 2

Identified Fe-containing phases and the best-fit spectral parameters as obtained for samples quenched into liquid nitrogen (T1): hyperfine field, *H*, average hyperfine field, *<H>*, both in kOe, isomer shift, *IS*, average isomer shift, *<IS>*, both in mm/s (relative to the source) and abundance, *A*, in percentage.

| Sample | CEMS | | | | | | | | | | | | | | TRANS | | |
|---|---|---|---|---|---|---|---|---|---|---|---|---|---|---|---|---|---|
| | Magnetite site A | | | Magnetite site B | | | Hematite | | | Fe-Cr | | | | Wüstite | | Fe-Cr | | |
| x | H | IS | A | H | IS | A | H | IS | A | $<H>$ | $<IS>$ | A | $x_1$ | $<IS>$ | A | $<H>$ | $<IS>$ | $x_2$ |
| 3.9 | 491.6 | 0.178 | 24.3 | 458.9 | 0.552 | 33.4 | 515.2 | 0.079 | 22.4 | 329.8 | -0.116 | 19.9 | 0.2(2) | - | - | 327.3 | -0.111 | 1.5(2) |
| 6.4 | 489.9 | 0.161 | 8.5 | 458.8 | 0.555 | 6.2 | 517.1 | 0.072 | 12.8 | 328.3 | -0.120 | 45.3 | 1.0(2) | 0.743 | 27.2 | 320.1 | -0.115 | 4.5(2) |
| 10.25 | - | - | - | - | - | - | - | - | - | 317.7 | -0.114 | 75.5 | 5.3(2) | 0.580 | 24.5 | 309.2 | -0.120 | 8.7(2) |
| 14.9 | - | - | - | - | - | - | - | - | - | 301.0 | -0.133 | 100 | 11.2(2) | - | - | 292.0 | -0.131 | 15.0(2) |

Fe-Cr alloys containing more than ~10 at% Cr are stainless. It has been theoretically explained as the consequence of complex competing magneto-chemical interactions between the alloying atoms [26,27]. On the other hand, all the spectra recorded in the transmission geometry show a pattern characteristic of a pure metallic (Fe-Cr) phase. This proves that only the pre surface zone had oxidized, while the bulk of the alloys had remained metallic. However, values of the spectral parameters of the metallic phase, and, in particular, those of the average hyperfine field, $<H>$, as obtained from the TRANS spectra have significantly lower values than those derived from the CEMS ones. The difference increases form 2.6 kOe for $x$=3.9 to 8.5 kOe for $x$=10.25. This indicates that the concentration of chromium in the metallic phase within the bulk is higher by ~1.5 at% for $x$=3.9 and by ~3.5 at% for $x$=10.25 as determined based on the $<H>$-$x$ relationship reported in Ref. 20, yet it is by ~2 at% lower than the initial values. Noteworthy, the decrease of $<H>$ could be also explained in terms of Cr atom clustering, however, such phenomenon for $Fe_{100-x}Cr_x$ alloys with $x \leq$~10 is not known to occur. Furthermore, as mentioned below for the *T2*-treated samples, an atom probe tomography analysis gave evidence in favor of the Cr depletion. Two conclusions can be drawn from these observations: first, the concentrations of chromium in the bulk of the *T1*-treated samples are lower than the original values which can be understood in terms of Cr atom segregation onto the surface which for $x \leq$10.25 was found to be multiphase. The metallic phase present in the pre surface zone is, however, richer in iron than the bulk one, what follows from the higher $<H>$-values found from the CEMS spectra. This means that the "missing" chromium is present in form of oxide phases, including chromium oxides that could not been revealed by the applied methods. The difference in the composition of both metallic phases relative to the initial concentration, $\Delta x$, is illustrated in Fig. 2. It is clear that the depletion of Cr in the metallic Fe-Cr phase as revealed from the CEMS spectra is in all samples higher than the one determined from the TRANS spectra. It is worth noting that the composition dependence of the degree of the Cr depletion is different: while a linear decrease is observed for the bulk, the depletion for the pre surface zone shows a maximum at $x \approx$8.

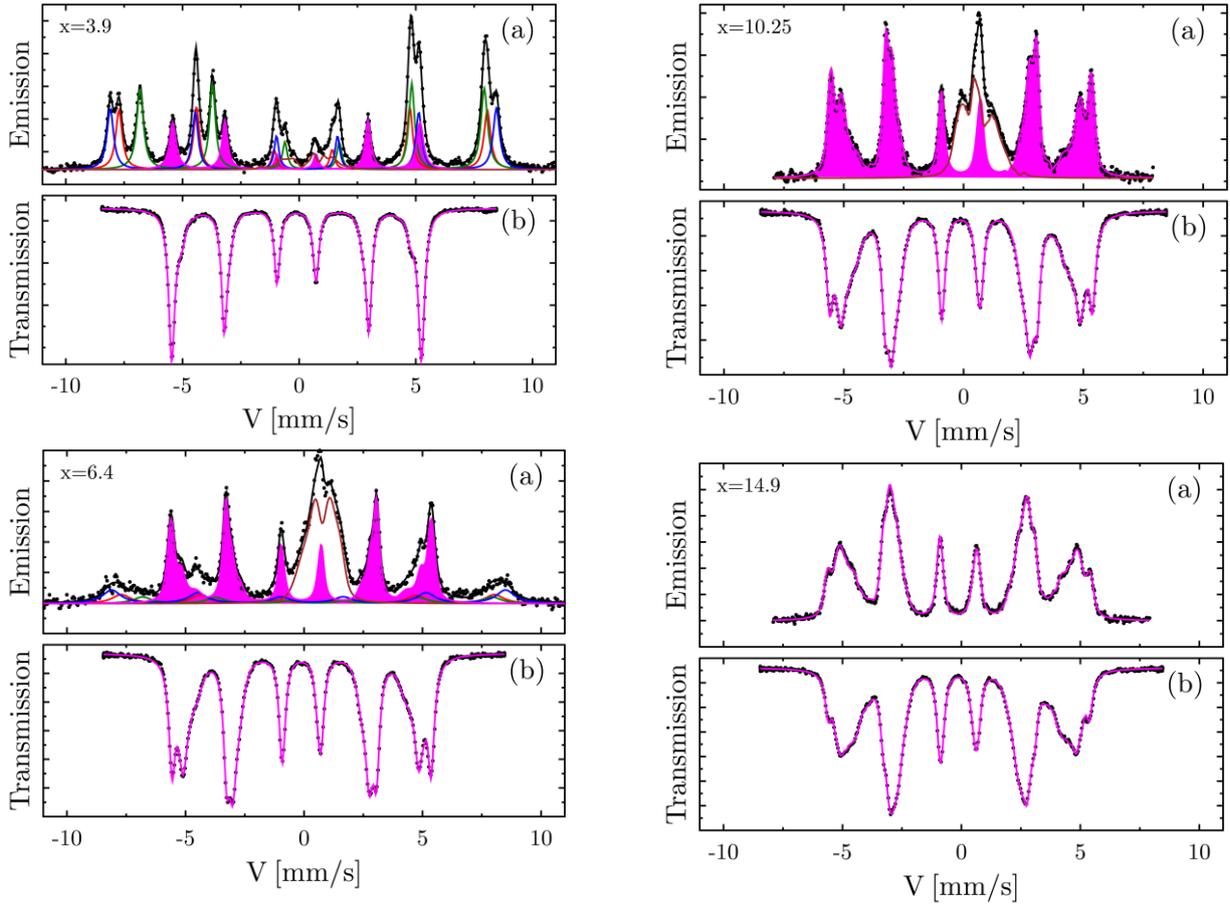

Fig. 1

*$^{57}$Fe spectra recorded on the Fe-Cr samples quenched into liquid nitrogen (T1). There are two spectra shown for each sample: CEMS spectrum (a) and TRANS spectrum (b). For the former the following Fe-containing phases are indicated: Fe-Cr or Fe (purple), magnetite (red and blue), hematite (green) and wüstite (brown).*

The enrichment of the surface of the Fe-rich Fe-Cr alloys in chromium is known to be responsible for their resistance to corrosion. However, how Cr atoms enrich the surface remains an open question. Some light on the issue was recently shed by performing Monte Carlo simulations [31], according to which this behavior was explained by a synergy between (1) the complex phase separation in the bulk alloy, (2) local phase transitions that tune the layers closest to the surface to an iron-rich state and inhibit the bulk phase separation in this region, and (3) its compensation by a strong and nonlinear enrichment in Cr of the next few layers. According to these calculations the enrichment increases strongly with Cr concentration reaching a narrow maximum at $x \approx 12$. Finally, it is worth noticing that the amount of the oxides as determined from the CEMS spectra is positively correlated with the observed decrease of Cr content in the bulk. As these oxides contain also chromium, this correlation can be taken as evidence for the segregation of Cr atoms into the surface-pre surface zone of the samples.

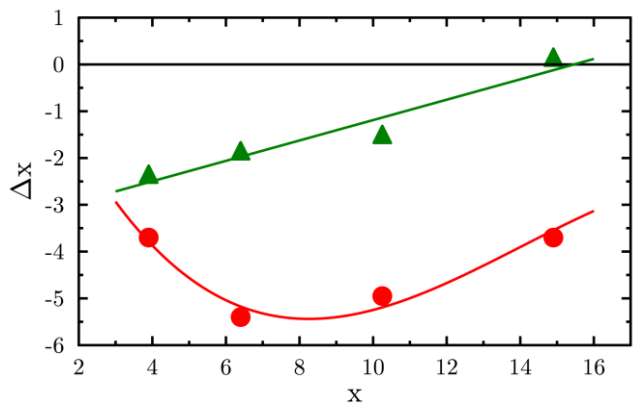

Fig. *2*

*Chromium depletion, Δx, in the metallic Fe-Cr phases of the pre surface zone (circles) and the bulk (triangles) versus the initial Cr content, x, for the T1-treated samples. Solid lines are guide to the eye.*

### 3.1. Treatment T2

Examples of the spectra recorded for the water-quenched samples are shown in Fig. 3 and the best-fit spectral parameters are presented in Table 3.

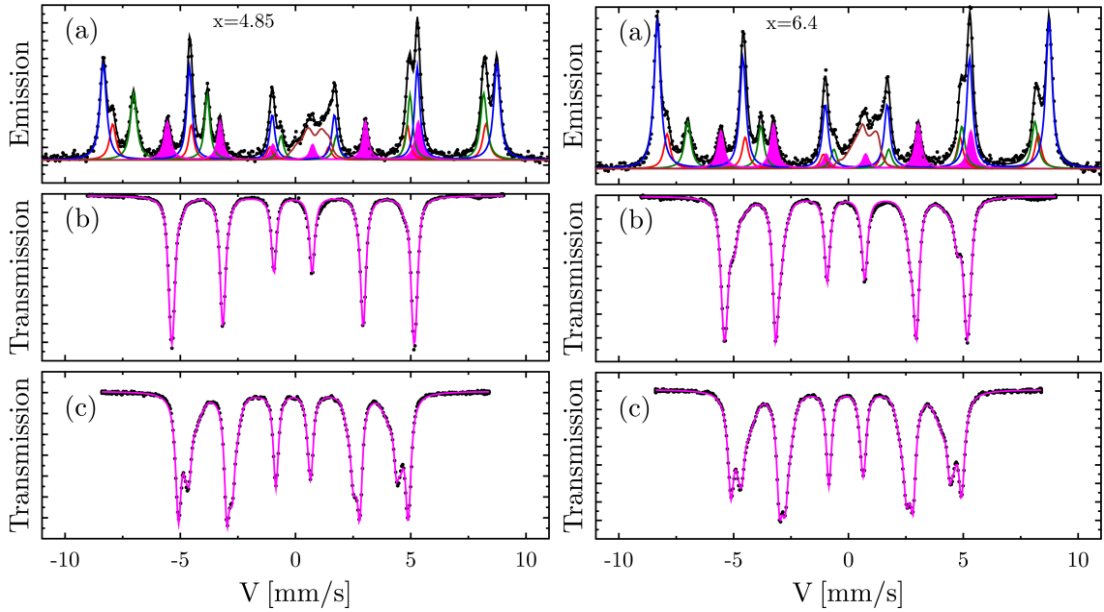

Fig. 3

*$^{57}$Fe spectra recorded on the Fe-Cr samples quenched into water. There are two spectra shown for each sample: CEMS spectrum (a) and TRANS spectrum (b). For the former the following Fe-containing phases are indicated: Fe-Cr or Fe (purple), magnetite (red and blue), hematite (green) and wüstite (brown). In addition, the corresponding spectra recorded for the untreated samples are added and labelled as (c).*

It can be easily noticed that in the case of the *T2*-treatment the difference between the spectra measured in the emission (CEMS) and in the transmission (TRANS) modes are even greater than that revealed for the *T1*-tratment. This feature is especially well visible when comparing the spectra measured in the transmission geometry for the $Fe_{95.15}Cr_{4.85}$ sample: the shape of the spectrum recorded in the TRANS mode on the quenched sample is almost the same as that of a pure iron what evidences a very high degree of chromium depletion in the bulk also. This effect was confirmed by the atom probe analysis with two different atom probes viz. laser and electric atom probes performed on the $Fe_{96.7}Cr_{3.3}$ sample viz. only Fe atoms were detected [32]. A selective removal of Cr atoms was already observed for Fe-Cr alloys that were sulphidized in an $H_2/H_2S$ atmosphere [33]. It is probably caused by a higher chemical affinity of sulphur/oxygen to chromium than to iron. The spectra recorded in the CEMS mode give evidence that the pre surface zone is composed of up to four phases, as specified in Table 3. However their relative abundance, $A$, strongly depends on the samples' composition. In particular, the amount of the Cr-substituted magnetite (seen by $^{57}$Fe atom probes) decreases from ~45% for $x$=2.2 at% to ~26% for $x$=6.4, hematite is present only in two Cr-least concentrated samples and its amount decreases from ~22% to ~13%, the amount of the metallic phase has a maximum at $x$=4.85 at%, and that of wüstite increases from ~8% for $x$=2.2 at% to ~15% for $x$=6.4at%. Again, comparing the spectral parameters of the magnetite shown in Table 3 with those reported in Ref. 30 one can conclude that the one present on the water-quenched samples can be identified as $Fe_{3-y}Cr_yO_4$ with $y\approx 0.5$ i.e. having a higher degree of Cr-substitution than that revealed in the case of the quenching into LN. The behavior of the abundance of the Fe-containing oxides observed with the increase of the chromium concentration is different than that revealed for the samples quenched into LN as it has a minimum at $x$=3.9.

Table 3
Identified Fe-containing phases and the best-fit spectral parameters as obtained for the pre surface zone of samples quenched into water: hyperfine field, $H$, average hyperfine field, $<H>$, both in kOe, isomer shift, $IS$, average isomer shift, $<IS>$, both in mm/s (relative to the source) and abundance, $A$, in percentage.

| Sample | Magnetite site A | | | Magnetite site B | | | Hematite | | | Fe-Cr | | | Wüstite | |
|---|---|---|---|---|---|---|---|---|---|---|---|---|---|---|
| x (at%) | H | IS | A | H | IS | A | H | IS | A | $<H>$ | $<IS>$ | A | $<IS>$ | A |
| 2.2 | 484.7 | 0.154 | 16.6 | 454.7 | 0.551 | 28.9 | 515.2 | 0.079 | 22.4 | 327.9 | -0.106 | 8.3 | 0.452 | 7.5 |
| 3.9 | 484.9 | 0.159 | 12.5 | 454.2 | 0.549 | 20.4 | 517.1 | 0.072 | 12.8 | 326.3 | -0.112 | 18.9 | 0.436 | 10.9 |
| 4.85 | 487.2 | 0.159 | 13.5 | 457.1 | 0.550 | 24.8 | - | - | - | 327.7 | -0.114 | 15.6 | 0.411 | 9.3 |
| 6.4 | 485.8 | 0.171 | 11.4 | 455.0 | 0.543 | 14.8 | - | - | - | 326.7 | -0.108 | 13.0 | 0.438 | 14.6 |

We notice that the average hyperfine field of the metallic phase hardly depends on the composition and ranges between 326.3 and 327.9 kOe. Based on the above-mentioned correlation between the average hyperfine field and Cr-concentration in Fe-Cr alloys, the Cr-content in the metallic phase present in the pre surface zone of the water-quenched samples was estimated to be in the range of 0.5-1 at%. Even more Cr-depleted turned out to be the bulk of the studied samples. In fact, the $<H>$-values derived from the TRANS spectra for $x$=2.2, 3.9 and 4.85 were characteristic of a pure iron, while that obtained for $x$=6.4, $<H>$=323.7 kOe, corresponds to $x$=3.4. This means that in all four samples quenched into water the depletion of chromium in the bulk took place. Taking into account the initial concentrations, the strongest depletion of ~4 at% occurred in the $Fe_{95.15}Cr_{4.85}$ sample.

### 3.1. Treatment T3

Figure 4 shows examples of the spectra recorded for the on-brass-quenched samples. There is a basic difference between the spectra recorded on the *T3*-treated samples and those recorded on both *T1*- and *T2*-treated ones. Namely, the CEMS spectra recorded on all *T3*-treated samples are characteristic of the metallic phase, Fe-Cr, only. In other words, the pre surface zone of the samples quenched onto the block of brass had not oxidized as in the case of LN- and water-quenched samples (chromium oxides may be present but they are not detectable with the applied techniques, as already mentioned). It is of interest to compare the spectral parameters of the corresponding spectra recorded in CEMS and TRANS modes, and, in particular, values of the average hyperfine field, $<H>$, as they can be used to evaluate the concentration of chromium in the pre surface zone, $x_1$, and in the bulk, $x_2$, respectively, using the $<H>$-$x$ correlation as reported elsewhere [20]. Such procedure can reveal possible differences in Cr atom distributions as obtained from the both types of the spectra. As displayed in Table 4, the $<H>$-values derived from the CEMS spectra are slightly higher than the corresponding ones obtained from the TRANS spectra for all $x$-values except for $x$=10.75 and 15.15, for which they are practically the same. In the light of the above-mentioned $<H>$-$x$ correlation it means that $x_1$ should be a little bit lower than $x_2$. In terms of the Cr depletion it means that the pre surface zone is for $x <$ ~10 slightly more depleted than the bulk. For the two most Cr-concentrated samples, the degree of depletion is similar which means that the pre surface zone of these samples does not differ from the bulk as far as the distribution of Cr atoms is concerned. In all cases, however, the *T3*-treatment resulted in a redistribution of Cr atoms that for $x <$ ~10 is different for the bulk than that for the pre surface zone. Figure 5 illustrates the estimated changes in the concentration of Cr relative to the initial values, in the pre surface zone, $\Delta x_1 = x_1 - x$, in the bulk, $\Delta x_2 = x_2 - x$, and between the pre surface zone and the bulk, $\Delta x_{12} = x_2 - x_1$.

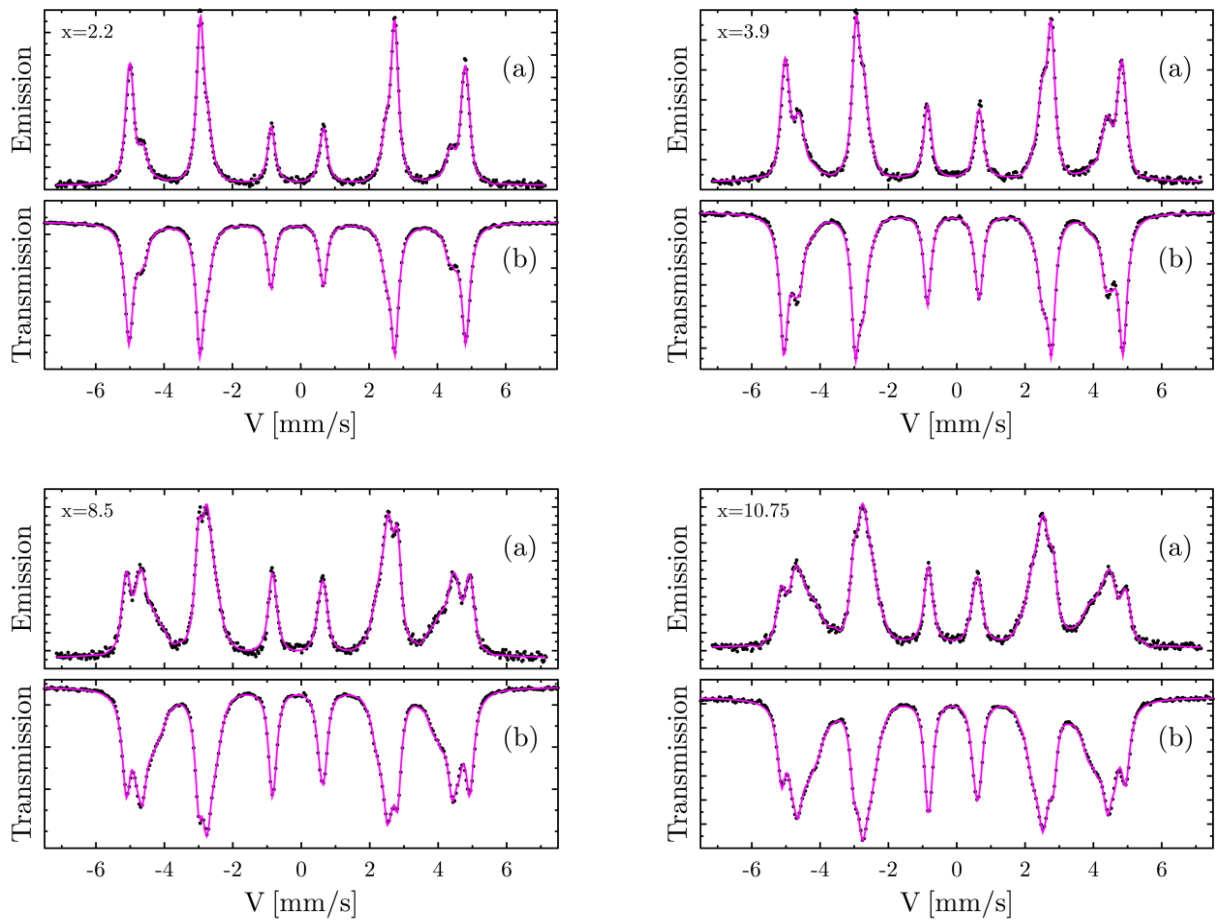

Fig. 4

*Examples of the $^{57}$Fe spectra recorded on the Fe$_{100-x}$Cr$_x$ samples quenched onto a block of brass. There are two spectra shown for each sample: CEMS spectrum (a) and TRANS spectrum (b).*

Table 4

The best-fit spectral parameters for the Fe$_{100-x}$Cr$_x$ samples quenched on the block of brass as obtained from emission (CEMS) and transmission (TRANS) spectra: average hyperfine field, $<H>$, in kOe, average isomer shift, $<IS>$, in mm/s (relative to the source) and abundance, $A$, in percentage. The Cr concentration derived from the $<H>$-values obtained from CEMS spectra is denoted by $x_1$ while that estimated from TRANS spectra by $x_2$ (both in at%).

| Sample | Fe-Cr CEMS | | | | Fe-Cr TRANS | | | |
|---|---|---|---|---|---|---|---|---|
| x (at%) | <H> | <IS> | A | $x_1$ | <H> | <IS> | A | $x_2$ |
| 2.2 | 327.9 | -0.108 | 100 | 0.5(2) | 325.7 | -0.112 | 100 | 1.8(2) |
| 3.9 | 323.7 | -0.110 | 100 | 2.6(2) | 322.3 | -0.113 | 100 | 3.1(2) |
| 8.5 | 313.4 | -0.121 | 100 | 5.5(2) | 311.6 | -0.112 | 100 | 6.3(2) |
| 10.75 | 305.1 | -0.127 | 100 | 8.0(2) | 304.2 | -0.122 | 100 | 8.2(2) |
| 15.15 | 292.1 | -0.133 | 100 | 11.6(2) | 292.4 | -0.133 | 100 | 11.5(2) |

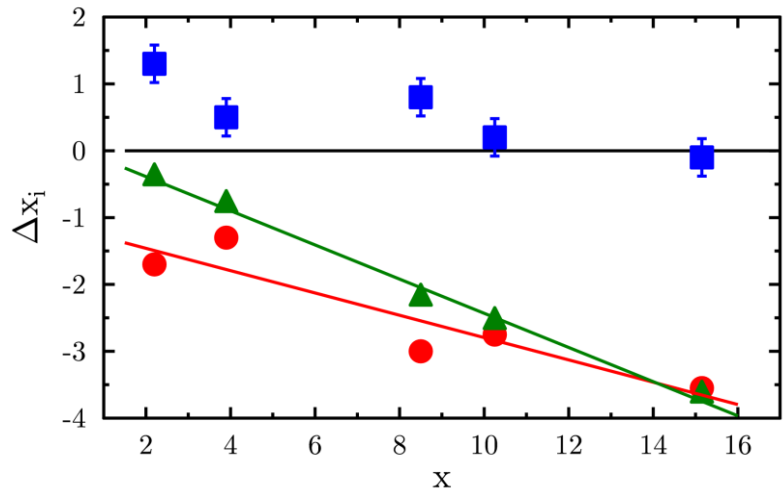

Fig. 5

*Chromium depletion, $\Delta x_i$, (i=1, 2, 21) as defined in the text, as found for the pre surface zone (circles), for the bulk (triangles) and the difference between the former and the latter (squares). Error bars are indicated for the latter. The lines are the best linear fits to the data.*

It is clear from Fig. 5 that the Cr depletion both in the pre surface zone as well as in the bulk increases with *x* which is opposite to the case of the *T1*-treated samples – see Fig. 2. The best linear fits to the data are indicated by full lines to highlight the trend. However, for *x* < ~10 the degree of the depletion is higher in the pre surface zone than that in the bulk, and it is the same for the two most Cr-concentrated samples. The difference $\Delta x_{21}=x_2-x_1$ is positive for *x* ≤~10 and equal to zero for higher *x*-values. Noteworthy, the finding reflects the known corrosion behavior of the Fe-Cr alloys and it shows that the stronger Cr depletion occurs in the concentration range where Fe-Cr alloys are prone to corrosion. This behavior is also evidenced by the results obtained on the T1/T2-treated samples as described above.

**3.2. SRO-parameters**

The SRO-parameters, $\langle\alpha_1\rangle$, $\langle\alpha_2\rangle$ and $\langle\alpha_{12}\rangle$, as determined for the T1- and T3-treated samples using equation (1) are displayed in Fig. 6.

It is clear that the random distribution of atoms was achieved neither with T1 nor T3 treatments. The quenching into LN (and water) has resulted not only in the oxidation of the pre surface zone of the samples with $x \leq \sim 11$, but also in a significant depletion of chromium in the bulk. Formally, the latter is reflected by high values the SRO parameters, especially for the low-concentrated samples. It is so because with the applied method we see only those Cr atoms that are present within the 1NN-2NN neighborhood of the probe $^{57}$Fe nuclei. Positive values of the SRO parameters mean that their actual number is less than expected for the random distribution. If the concentration of Cr were constant the positive SRO-parameters values would indicate clustering of Cr atoms. As this is not the case, one has to conclude that Cr atoms had segregated onto the surface and the positive values of the SRO-parameters should be interpreted as indicative of the Cr depletion in the bulk.

The samples that underwent the quenching onto the block of brass have not been oxidized (at least as far as Fe-containing oxides are concerned) and the distribution of atoms is fairly random, especially as far as $\langle\alpha_{12}\rangle$ is concerned. Here, a deviation from the randomness towards Cr atoms clustering occurs only for $x \geq \sim 15$. However, the behavior observed for particular co-ordinations shells is different and in both cases a deviation from the randomness, albeit in the opposite direction – with an inversion at $x \approx 5$ – is observed. This interpretation is correct assuming a lack of Cr depletion in the bulk. What had really happened with Cr atoms due to the applied treatments cannot be answered with the presently applied methods.

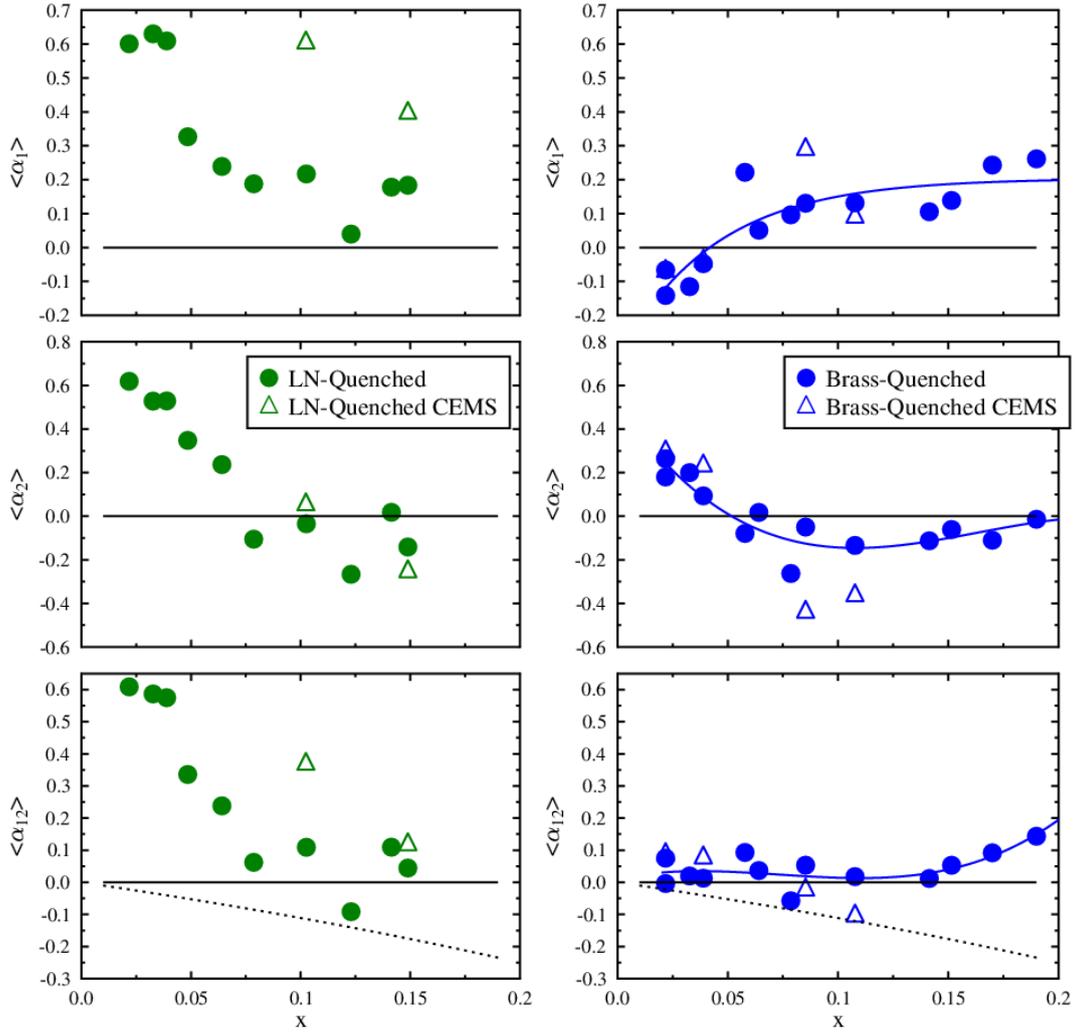

Fig. 6

*The SRO-parameters $<\alpha_1>$, $<\alpha_2>$ and $<\alpha_{12}>$ versus Cr content, x, for the $Fe_{100-x}Cr_x$ samples quenched into LN (left panel) and onto a brass block (right panel). Circles represent the data obtained from the spectra measured in the transmission mode, while triangles stand for those derived from the CEMS spectra. The dotted line shows the theoretically allowed limit.*

## 5. Conclusions

The results obtained in this study gave a clear evidence that a quenching medium is of importance as far as a distribution of Cr atoms in Fe-rich Fe-Cr alloys is concerned. In particular, the following conclusions can be drawn:

1. Quenching into water and liquid nitrogen resulted in a formation of up to three different Fe-containing oxides (magnetite, hematite and wüstite) on a surface while samples quenched onto a block of brass were free of such oxides.

2. Short-range order parameters for the bulk metallic Fe-Cr phase are characteristic of a heat treatment and for a given treatment they are typical of the coordination shell (1NN, 2NN).

3. Quenching onto the block of brass resulted in a quasi-random distribution of Cr atoms within the 1NN-2NN volume for $x<\sim15$at%Cr, and clustering for higher $x$-values. The distribution in the individual neighbor shells i.e. 1NN and 2NN was, however, not random with an inversion in $<\alpha_1>$ and $<\alpha_2>$ at $x\approx5$ albeit going in the opposite direction.

4. Quenching into all three media resulted in a depletion of Cr atoms in the bulk of the samples, but for LN and $H_2O$ the degree of the depletion was decreasing while for the brass it was increasing with $x$.

**Acknowledgements**

This study was carried out within the IPPLM-EURATOM Association. It was also supported by The Ministry of Science and Higher Education, Warszawa, Poland. Cristelle Pareige is thanked for performing APT analysis.